\documentclass[s12pt]{iopart}
\usepackage{graphicx}
\usepackage[utf8]{inputenc}
\usepackage{verbatim}
\usepackage{amssymb}
\usepackage{amstext}
\usepackage{listings}
\usepackage{color}
\usepackage{hyperref}
\usepackage{color}
\usepackage{xspace}

\newcommand\subfig[2]{{figure~\ref{#1}{#2}}}
\newcommand\subcap[1]{{(#1):}}
\usepackage{bold-extra}

\newcommand{\DA}{\theta} 
\newcommand{\DD}{\rho} 
\newcommand{\RDA}{\tilde{\theta}} 
\newcommand{\DIR}{\phi} 
\newcommand{\RDIR}{\phitilde} 

\newcommand{\CHDIR}{\Delta\phi} 
\newcommand{\TS}{t} 
\newcommand{\RAT}{\eta} 
 
\newcommand{\LINE}{\lvec} 
\newcommand{\LCUTP}{\lambda^+_{\text{cut}}} 
\newcommand{\LCUTM}{\lambda^-_{\text{cut}}} 
\newcommand{\LCUTPM}{\lambda^{\pm}_{\text{cut}}} 

\newcommand{\SEG}{\SCAL} 
\newcommand{\IMPACT}{impact parameter\xspace} 
\newcommand{\TMIX}{t_{\text{mix}}}
\newcommand{\ZIG}{ZZ}

\newcommand{\SET}[1]{\{#1\}}

\newcommand{\quot}[1]{``#1''}

\newcommand{\DCAL}{\mathcal{D}}  
\newcommand{\OCAL}{\mathcal{O}}  
\newcommand{\SCAL}{\mathcal{S}}  
\newcommand{\sina}[2][]{\sin^{#1} \! \gla #2 \gra}  
\newcommand{\cosa}[2][]{\cos^{#1} \! \gla #2 \gra}  

\newcommand{\sinb}[2][]{\sin^{#1} \glb #2 \grb}  

\newcommand{\expecta}[1]{\mathbb{E} \gla #1 \gra}

\newcommand{\expectb}[1]{\mathbb{E} \glb #1 \grb}

\newcommand{\gla}{\,}  
\newcommand{\gra}{}  
\newcommand{\glb}{\left(}  
\newcommand{\grb}{\right)}  
\newcommand{\glc}{\left[}  
\newcommand{\grc}{\right]}  

\newcommand{\const}{\text{const}}

\newcommand{\VEC}[1]{\mathbf{#1}}
\newcommand{\lvec}{\VEC{l}}

\newcommand{\xvec}{\VEC{x}}
\newcommand{\xvecbar}{\overline{\VEC{x}}}

\newcommand{\phitilde}{\tilde{\phi}}

\newcommand{\half}{\frac{1}{2}}

\newcommand\diff[1]{\mathrm{d}#1}

\newcommand\bigO[1]{\ensuremath{\OCAL(#1)}}

\newcommand{\figurescale}{1.1}

\begin{document}
\date{\today}
\title{Direction-sweep Markov chains}

\author{Liang~Qin$^1$, Philipp~Höllmer$^2$ and Werner~Krauth$^1$ }

\address{$^1$ Laboratoire de Physique de l’Ecole normale supérieure, ENS, 
Université PSL, CNRS, Sorbonne Université, Université de Paris,
Paris, France}
\address{$^2$ Bethe Center for Theoretical Physics, University of Bonn, 
Nussallee 12, 53115 Bonn, Germany}
\ead{\mailto{liang.qin@phys.ens.fr}, \mailto{hoellmer@physik.uni-bonn.de}, 
\mailto{werner.krauth@ens.fr}}


\begin{abstract}
We discuss a non-reversible, lifted Markov-chain Monte Carlo (MCMC) algorithm 
for particle systems in which the direction of proposed displacements is changed 
deterministically. This algorithm sweeps through directions analogously to the 
popular MCMC sweep methods for particle or spin indices. Direction-sweep MCMC 
can be applied to a wide range of original reversible or non-reversible Markov 
chains, such as the Metropolis algorithm or the event-chain Monte Carlo 
algorithm. For a single two-dimensional dipole, we consider direction-sweep MCMC 
in the limit where restricted equilibrium is reached among the accessible 
configurations before changing the direction. We show rigorously that 
direction-sweep MCMC leaves the stationary probability distribution unchanged, 
and that it profoundly modifies the Markov-chain trajectory. Long excursions, 
with persistent rotation in one direction, alternate with long sequences of 
rapid zigzags  resulting in persistent rotation in the opposite direction in the 
limit of small direction increments. The mapping to a Langevin equation then 
yields the exact scaling of excursions while the zigzags are described through a 
non-linear differential equation that is solved exactly. We show that the 
direction-sweep algorithm can have shorter mixing times than the algorithms with 
random updates of directions. We point out possible applications of 
direction-sweep MCMC in polymer physics and in molecular simulation.
\end{abstract}

\vspace{28pt plus 10pt minus 18pt}
\noindent{\small\rm 
This is the version of the article before peer review or editing, as submitted 
by an author to \jpa IOP Publishing Ltd is not responsible 
for any errors or omissions in this version of the manuscript or any version 
derived from it. The Version of Record is available online at 
\url{https://doi.org/10.1088/1751-8121/ac508a}.
\par}

\maketitle

\section{Introduction}

Since its introduction in $1953$, the Markov-chain Monte Carlo (MCMC) 
method~\cite{Metropolis1953JCP} has developed into an essential tool in science 
and engineering, and into a prominent mathematical research 
discipline~\cite{Levin2008}. 
MCMC is concerned with the sampling of a probability distribution $\pi$,
for example a
Boltzmann distribution in equilibrium statistical physics.
MCMC's trademark 
properties are randomness and absence of memory: Samples $j$ at Monte-Carlo 
time 
step $\TS + 1$ are produced from samples $i$ at time step $\TS$ with 
independent 
probabilities contained in a transition matrix $P = (P_{ij})$. The stationary 
distribution $\pi$ is reached in the limit $\TS \to \infty$. It satisfies 
the global-balance condition $\sum_i \pi_i P_{ij} = \pi_j$. Most MCMC 
algorithms 
are reversible. They satisfy the detailed-balance condition $\pi_i P_{ij} =  
\pi_j P_{ji}$ that implies the less restrictive global balance by summing over 
$i$. In physical terms, a reversible Markov chain implements equilibrium 
dynamics that approaches the Boltzmann distribution, with the detailed-balance 
condition expressing the vanishing of all flows. In contrast, 
a non-reversible Markov chain implements 
out-of-equilibrium dynamics with a steady state (imposed by the global-balance 
condition) that again coincides with the Boltzmann distribution $\pi$.

Under the necessary conditions of 
irreducibility and aperiodicity~\cite{Levin2008}, MCMC algorithms often allow 
for sequential variants that 
seemingly introduce memory effects to the move (the move at time $t+1$ may 
depend on the move at time 
$t$).
For systems of $N$ particles, one such variant was pioneered in the original 
1953 
reference: Instead of attempting at time $t$ a move
of a randomly chosen particle
\quot{\dots we move each 
of the particles in succession \dots}~\cite[p.22]{Metropolis1953JCP}, that is, 
attempt a  move of particle $i+1$ (modulo $N$)  at time $t+1$ after an attempt 
of $i$ at time $t$. 
In particle systems with central potentials, this particle-sweep
Monte Carlo algorithm is somewhat faster (apparently by a constant factor)
than the 
random-choice variant~\cite{OKeeffe2009,KapferKrauth2017,Lei2018b, 
RenOkeeffe2007}. 
In the Ising model and related systems, sequential  updates of spin $i+1$ after 
spin $i$, etc., (\quot{spin sweeps}) also break detailed balance yet 
satisfy global balance. Spin-sweep algorithms are again faster, by a constant 
factor, than the detailed-balance 
MCMC that flip spins in random order~\cite{Berg2004book}.

Lifting~\cite{Diaconis2000} allows one to  formulate such a partly 
deterministic algorithm 
as a Markov chain with a time-independent transition matrix, and to 
expose 
its close connection with the \quot{collapsed} Markov chain in which 
moves are proposed randomly. The lifted transition matrix 
acts on lifted (extended) samples. In the above example, particle-lifted 
samples comprise the active-particle 
index in addition to all the particle coordinates. 
The particle- and spin-sweep algorithms 
are 
precursors of non-reversible Markov chains that have been much studied in the 
recent past~\cite{SuwaTodoPRL2010,Turitsyn2011,FernandesWeigelCPC2011,
BierkensPDMC2017,Bierkens2017}. One powerful 
non-reversible 
MCMC method is the event-chain Monte Carlo 
(ECMC) algorithm~\cite{Bernard2009,Michel2014JCP,Krauth2021eventchain}.

In more than one dimension, naturally, particles must move into different 
spatial
directions. Rather than to sample the 
direction of the proposed move at time $t$ randomly, one may define it as a 
lifting variable, and modify it deterministically without influencing the 
stationary distribution. In the family of ECMC algorithms, direction lifting 
can be 
applied to straight ECMC that uses the same direction for every move in an 
event chain (in contrast to, e.g., Newtonian~\cite{Klement2019} and 
Forward ECMC~\cite{Michel2020} that modify the direction in each event).
Very slow changes of the direction after each event chain
were studied for straight ECMC simulations of hard-sphere systems, where they 
were 
not found to improve the convergence properties~\cite{Weigel2018}.

In the present paper, we discuss direction-sweep MCMC for a simplified
two-dimensional model of an extended flexible dipole consisting of two atoms, 
that resembles a flexible water molecule in the context of molecular simulation 
(see~\cite{Faulkner2018, Hoellmer2020}). The
variable dipole length allows the entire dipole to rotate through local
MCMC moves of both atoms along slowly changing directions. 
We obtain exact results for direction sweeps in the
limit where a given direction remains fixed until restricted equilibrium is
reached among the accessible configurations by the application of any local MCMC
algorithm (whose moves can be constructed with finite acceptance probability
from a sequence of infinitesimal moves). We show analytically that
slow direction sweeps (small direction-angle increments) yield long-lived 
rotations of the dipole by a cumulative (\quot{rolled-out}) mean 
rotation angle that diverges as the inverse angle increment in the limit of 
infinitely slow sweeps. This 
motion is interrupted by a
counter-rotation that proceeds in a fast sequence of steps. Both motions exactly
balance, and the expectation of the net rotation vanishes identically.
Numerically, we show that direction sweeps can lower mixing times
compared to MCMC with random choice of the direction. We
find that only picking directions among the $x$- and $y$-axes is by far the most
unfavorable case, although it was previously used in applications of straight 
ECMC to
dipolar systems~\cite{Faulkner2018, Hoellmer2020}. Our results, for dipoles,  
differ from those for hard-sphere systems and probably, more generally, from 
those for
simple liquids~\cite{Weigel2018}.

Our simple dipole model serves as an analytically tractable test bed for
molecules such as the simple-point-charge-with-flexible-water 
(SPC/Fw) water model~\cite{WuTepperVoth2006}. The model is also very closely 
related to the flexible polymer models that are being intensely studied using 
ECMC~\cite{Mueller2020,Kampmann2021}.
Direction lifting is insensitive to the dipole's structure and 
interactions. It remains valid for $N$-particle systems. 

The paper is organized as follows: \Sref{sec:model} introduces the dipole 
model. In \sref{sec:sequential}, we introduce direction-sweep MCMC  
that reaches 
restricted equilibrium in a single step, and prove that it
converges towards the stationary distribution $\pi$. 
\Sref{sec:equilibrium} studies the trajectory of the dipole orientation. Mixing 
times are determined and compared in \sref{sec:convergence}. 
\Sref{sec:conclusion} provides a summary of our main results.

\section{Dipole Model}
\label{sec:model}

\begin{figure}
\begin{indented}
\item[]
\includegraphics[scale=\figurescale]{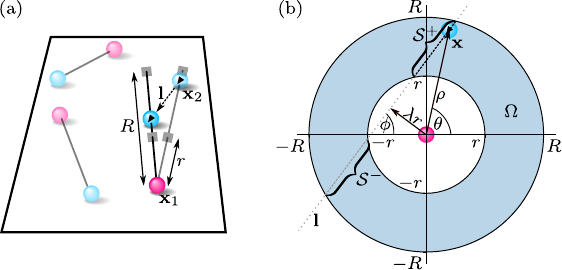}
\end{indented}
\caption{Two-dimensional dipole of length $\DD\in[r,R]$. \subcap{a} MCMC move 
of 
a dipole atom along the line $\LINE$. \subcap{b} The same move in the 
ring representation $ \xvec_2 - \xvec_1 = (\DD, \DA) \in \Omega$.
The line $\LINE$  has angle 
$\DIR$ with the $x$-axis and distance $\lambda r$ to the origin. For 
$|\lambda|<1$, it forms two segments $\SEG^-, \SEG^+ \subset
\LINE \cap \Omega$.}

\label{fig:RingRepresentation}
\end{figure}

We study a two-dimensional flexible dipole consisting of two 
atoms $1$ and $2$ with an interaction that only enforces a minimum length $r$ 
and a maximum length $R$. Specifically, the two atoms are at positions $\xvec_1 
$ and $\xvec_2$ in a two-dimensional homogeneous domain. The flat interaction
\begin{equation}
U(\DD)= \cases{
0 & if $\DD \in [r,  R]$,\\
\infty & otherwise,}
\label{equ:InnerDipolePotential}
\end{equation}
only depends on the dipole length $\DD = |\xvec_2 - \xvec_1|$. The 
model can be generalized to three spatial dimensions, and also made more 
realistic, for example through the 
SPC/Fw water model~\cite{WuTepperVoth2006}. The single dipole of 
\eref{equ:InnerDipolePotential} is to be envisaged as part of a more complex 
many-dipole system with hard-sphere atomic pair interactions, that we will 
however not study in the present paper (see \subfig{fig:RingRepresentation}{a}).

Any local MCMC move proposes a displacement of $\xvec_1$ or $\xvec_2$ from its 
present position along a line $\LINE$ of angle $\DIR$ with the $x$-axis. If the 
final configuration has a dipole length with infinite $U$, the move is 
rejected. Such single-atom moves induce translations and rotations of 
the entire dipole. We need not consider explicit global MCMC 
translations or global rotations which, in the application to ECMC that we have 
in mind, are difficult to implement.
Because of homogeneity, uniform translations of the dipole 
decouple from its rotations. We may thus set $ \xvec_1 = (0,0)$ and consider the 
relative position in polar coordinates: $\xvec= \xvec_2 - \xvec_1 = (\DD, 
\DA)$, with the dipole angle $\DA$. In equilibrium, 
$\xvec$ is uniformly distributed on the sample space $\Omega$, 
the two-dimensional ring of inner 
radius $r$ and outer radius $R$ centered at $(0,0)$ (see 
\subfig{fig:RingRepresentation}{b}). The uniform Euclidean 
distribution on $\Omega$ translates into a dipole-length distribution $\pi(\DD)= 
2 \DD/ \glc r^2\glb \RAT^2 - 1\grb \grc$ for $\DD \in [r, R]$ with 
$\RAT=R/r$, and a dipole-angle distribution for $\DA$ that is uniform 
in $(-\pi,\pi]$.

The sampling of the dipole $(\xvec_1, \xvec_2)$ may be tracked in the ring 
representation for $\xvec \in \Omega$, because any move of $\xvec_2$ corresponds 
to the identical move for $\xvec$, and any move of $\xvec_1$ yields an inverse 
move for $\xvec$. In both cases, the move is on a line $\LINE \ni 
\xvec$ of angle $\DIR$ with the $x$-axis. We thus
parameterize a direction of a local MCMC move by $\DIR$. The 
trajectory of the dipole angle under slow direction sweeps is closely 
connected to the trajectory of the impact parameter 
\begin{equation} 
\label{eq:impact}
\lambda = \frac{\sin(\DA - \DIR)\DD}{r}, 
\end{equation} 
which denotes the signed distance (in units of $r$) of $\LINE$ to the origin in 
a local MCMC move. In the 
reference frame where $\DIR=0$, 
i.e., where $\LINE$ runs parallel to the $x$-axis, $\lambda$ is positive for 
$\xvec$ in the upper half plane and negative in the lower half plane. If 
$|\lambda| > 1$, $\LINE \cap \Omega$ forms a single segment $\SEG$  that 
contains all 
accessible configurations. If 
$|\lambda| < 1$, $\LINE \cap \Omega$ forms two such segments, namely $\SEG^-$ 
(to the 
left for $\DIR=0$) and $\SEG^+$ (to the right) (see 
\subfig{fig:RingRepresentation}{b}). In realistic applications like, e.g., 
dense dipole systems, accepted local MCMC moves with $|\lambda| < 1$ that jump 
from $\SEG^-$ to $\SEG^+$ are highly unlikely. We thus only 
consider local MCMC 
moves that remain within their respective segment ($\SEG$, $\SEG^-$ 
or $\SEG^+$) or, in other words, local MCMC moves that can be constructed from 
infinitesimal legal moves.

\section{Direction-sweep MCMC}
\label{sec:sequential}

\begin{figure}
\begin{indented}
\item[]
\includegraphics[scale=\figurescale]{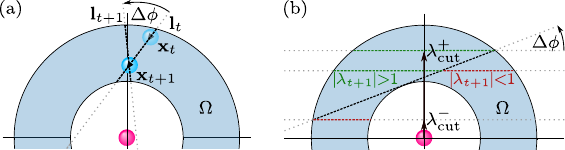}
\end{indented}
\caption{Direction-sweep MCMC with directions $\DIR_{t+1} - \DIR_t = 
      \CHDIR$. \subcap{a} Lines $\LINE_\TS \ni \xvec_t$ and 
      $\LINE_{\TS + 1}$ intersect at $\xvec_{\TS + 1}$. 
      \subcap{b} A large impact parameter ($|\lambda_\TS| > \LCUTP$) leads to  
      $|\lambda_{\TS+1}|> 1 \quad  \forall \xvec_{\TS + 1} \in \LINE_\TS$. 
      Likewise, a small impact parameter ($|\lambda_\TS| < \LCUTM$) leads to 
      $|\lambda_{\TS + 1}| < 1 \quad \forall \xvec_{\TS + 1} \in \LINE_\TS$.
      Otherwise, if $\LCUTM < |\lambda_\TS| < \LCUTP$, the value of $\xvec_{\TS 
      + 1}$ 
      determines whether $|\lambda_{\TS + 1}| > 1$ or $|\lambda_{\TS+1}| < 1$.}
\label{fig:Sequential}
\end{figure}

Local MCMC moves along one direction $\DIR$ tend towards a 
restricted equilibrium  among the accessible configurations in $\SEG$, $\SEG^-$ 
or $\SEG^+$.
For the single two-dimensional dipole, the restricted-equilibrium 
limit can be reached in a
single step by sampling the final position of the dipole 
in the
segment that contains the starting position $\xvec$. This allows us to focus on
the effects introduced by the choice of directions. In the following, one unit
of MCMC time corresponds to one fixed direction. We therefore 
obtain the
next position $\xvec_{\TS + 1}$ as a direct uniform sample~\cite{SMAC} in 
$\SEG$ for $|\lambda_\TS| > 1$, and in $\SEG^-$ or in $\SEG^+$ (depending 
on the starting position $\xvec_\TS$) for $|\lambda_\TS| < 1$. The direction is 
incremented by a constant value after each time step, that is, the line 
$\LINE_{\TS + 1}$ goes through $\xvec_{\TS + 1}$ with the new angle 
$\DIR_{\TS+1} =  \DIR_{\TS} + \CHDIR$ (for concreteness, we suppose $\CHDIR > 
0$) (see \subfig{fig:Sequential}{a}). 
The original (\quot{collapsed}) 
reversible version of direction-sweep MCMC randomly samples the new angle from 
$\DIR_{\TS+1} = \DIR_{\TS} \pm \CHDIR$ instead. In this case, detailed balance 
follows from the fact that both the choice of $\xvec_{\TS+1}$ from 
$\xvec_\TS$, 
and of $\DIR_{\TS+1}$ from $\DIR_\TS$ are reversible.

Because of the $\pi$-periodicity of the directions, a given value of $\CHDIR$ 
and the starting direction $\DIR_0$ imply a set of directions 
$\DCAL=\{\DIR_\TS\}$ that contains all possible directions of a simulation. We 
only consider finite direction sets $\DCAL$ for simplicity. 
Direction-sweep MCMC is a non-reversible lifting of reversible local MCMC
with an 
augmented (lifted) sample space $\Omega \times \DCAL$. The lifted stationary 
distribution
depends on $\DIR$: $\pi(\xvec) \mapsto
\pi(\xvec,\DIR)$, with $\sum_\DIR \pi(\xvec,\DIR) = \pi(\xvec)$ 
(see \cite{Chen1999,Krauth2021eventchain} for definitions). We will show 
below that 
$\pi(\xvec, \DIR)$ is proportional to $\pi(\xvec)$ for the sequential direction 
sweep. In order to converge towards the stationary distribution $\pi$, the 
direction-sweep MCMC algorithm must satisfy the global-balance condition: 
\begin{equation}
   \pi(\xvec_{\TS+1},\DIR_{\TS+1}) = \sum_{\xvec_\TS,\DIR_\TS}
   \pi(\xvec_\TS, \DIR_\TS) \,
   p[(\xvec_\TS, \DIR_\TS) \rightarrow (\xvec_{\TS+1}, \DIR_{\TS+1})],
\label{eq:global}
\end{equation}
where $p(a \rightarrow b)$ denotes the transition probability between the 
(lifted) configurations $a$ and $b$. 

As explained, any move from $t$ to $t+1$ 
is composed of two parts. In the first part, the lifting variable is 
fixed ($\DIR_{t+1} = \DIR_t=\DIR$) while a new position $\xvec_{\TS+1}$ is 
directly sampled among the accessible configurations given $\xvec_\TS$ and 
$\DIR_\TS$. This restricted MCMC algorithm satisfies global balance for the 
given direction $\DIR$ by construction, that is, $\pi(\xvec_{\TS + 1}, \DIR) = 
\sum_{\xvec_\TS} \pi(\xvec_{\TS}, \DIR)\,p[(\xvec_{\TS}, \DIR) \rightarrow 
(\xvec_{\TS + 1}, \DIR)]$. This yields specifically $\pi(\xvec_{\TS+1}, \DIR) = 
\pi(\xvec_{\TS}, \DIR)$ for the simple dipole model with its uniform $\pi$. In 
the second part of the move, the position $\xvec$ 
remains fixed and only the direction is incremented by 
$\CHDIR$:
\begin{equation}
   \pi(\xvec,\DIR_{\TS+1}) = \sum_{\DIR_\TS} 
   \pi(\xvec, \DIR_\TS) \,
   p[\DIR_\TS  \to \DIR_{\TS+1}] = \pi(\xvec, \phi_{\TS+1} - \CHDIR). 
\label{eq:globalTwo}
\end{equation}
For a finite direction set, and taking into account the periodicity of 
directions, this establishes that the lifted stationary 
distribution $\pi(\xvec, 
\phi)$ is independent of $\DIR$. The first and second parts together establish 
the validity of the global-balance condition in \eref{eq:global}. For the 
two-dimensional dipole model, direction-sweep MCMC is aperiodic and irreducible 
for any choice of two or more directions. The independence of the stationary 
distribution $\pi(\xvec, \phi)$ with respect to  the lifting variable is a 
general property of lifted MCMC \cite{Krauth2021eventchain}.

For small $\CHDIR$, direction-sweep MCMC 
features two cutoff impact parameters $\LCUTP$ and $\LCUTM$:
\begin{equation}
	\LCUTPM = \cos(\CHDIR) \pm \sqrt{(\RAT^2 - 1)\sin^2(\CHDIR)}.
	\label{eq:lcut}
\end{equation}
Here, $\LCUTP$ exists for $\cos(\CHDIR) > 1 / \RAT$, and $\LCUTM$ for
$\cos(\CHDIR) > (\RAT^2 - 2) / \RAT^2$. For a fixed value of $\RAT$, $\LCUTPM$ 
both exist in the limit $\CHDIR \to 1$ and approach $1$ as $\LCUTPM \to 
1^{\pm}$. The interval $[\LCUTM, \LCUTP]$ is a 
separation layer for the impact parameter $\lambda$ because
$|\lambda_t| > \LCUTP$ implies $|\lambda_{\TS+1}|  > 1$, whereas
$| \lambda_t|  < \LCUTM$ implies $| \lambda_{\TS+1}|  < 1$ (see 
\subfig{fig:Sequential}{b}). 

\section{Equilibrium Properties}
\label{sec:equilibrium}

\begin{figure}
\begin{indented}
\item[]
\includegraphics[scale=\figurescale]{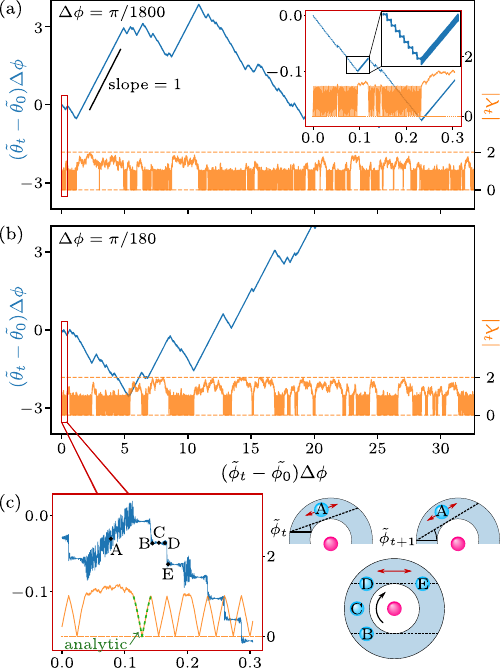}
\end{indented}
\caption{Trajectories of $(\RDA_\TS - \RDA_0)\CHDIR$ (blue, upper) 
      and of $|\lambda_\TS|$ (yellow, lower) \emph{vs.}
      $(\RDIR_\TS - \RDIR_0) \CHDIR $ for $\RAT = 2$.
      \subcap{a} $\CHDIR = \pi / 1800$. The inset shows the initial 
      trajectories. \subcap{b} $\CHDIR=\pi/180$. 
      \subcap{c} Initial trajectories for $\CHDIR=\pi / 180$, 
      with \quot{excursions} (upper right) and 
      \quot{zigzags} (lower right).
}
\label{fig:ThetaLambdaTrajectory}
\end{figure}

For small $\CHDIR > 0$, direction-sweep MCMC simulations of the 
single 
two-dimensional dipole yield trajectories of $\RDA_{\TS}$, the rolled-out 
dipole angle (not wrapped back into a $2 \pi$ interval), with alternating 
positive and negative rotations.
The positive rotations fluctuate around their 
average of $\CHDIR$ per time step, so that the 
trajectory of $(\RDA_{\TS} - \RDA_0) \CHDIR$ \emph{vs}
$(\RDIR_{\TS} - \RDIR_0) \CHDIR$ has an average unit slope 
($\RDIR_{\TS}$ is the rolled-out 
direction: $\RDIR_{\TS} - \RDIR_0 = t\CHDIR$). 
The negative rotations exhibit, in contrast, 
intermittent sharp 
decreasing steps and constant plateaus (see \fref{fig:ThetaLambdaTrajectory}). 

We will show that the trajectory of $\RDA_t$ depends on that of the impact 
parameter $\lambda_\TS$ (see sections~\ref{sec:excursions} 
and~\ref{sec:zigzags}). Therefore, 
we first treat the 
trajectory of $\lambda_\TS$ that may be described through 
drift and diffusion terms. 
The current value of the impact parameter $\lambda_{\TS}$ is a function of 
the position $\xvec_\TS$ and
direction $\DIR_\TS$ (for fixed system parameters).
Direction-sweep MCMC algorithm then uniformly samples 
$\xvec_{\TS + 1}$ on the segment  that also contains $\xvec_\TS$. 
The subsequent increment of the direction by $\CHDIR$ linearly maps 
$\xvec_{\TS + 1}$ onto the impact parameter $\lambda_{\TS + 1}$. The position
$\xvec_{\TS + 1}$, given $\xvec_{\TS}$, is a random variable, and so is 
$\lambda_{\TS + 1}$. Its conditional expectation is
\begin{equation}
\expecta(\lambda _{\TS+1}|\lambda _{\TS}) = \cases{
\lambda_\TS \cosa{\CHDIR}  \mp A_\TS \sina{\CHDIR} & if 
$|\lambda_{\TS}| < 1$; $\xvec_t \in \SEG^\pm$, \\
\lambda_\TS \cosa{\CHDIR} \hphantom{\mp A_\TS \sina{\CHDIR}s} & if 
$|\lambda_\TS| > 1$,
}
\label{equ:Drift}
\end{equation}
where $A_\TS = ( B_\TS + C_\TS ) / 2$ with $B_\TS=\sqrt{\RAT^2 - 
\lambda_{\TS}^2}$ and $C_\TS = \sqrt{1 - \lambda_{\TS}^2}$. The variance of 
$\lambda_{\TS+1}$ is
\begin{equation}
 \sigma^2(\lambda_{\TS + 1}|\lambda _{\TS})  = \cases{
 \frac{1}{12} \sinb[2]{\CHDIR} \glb B_\TS - C_\TS \grb^2 & if 
 $|\lambda_{\TS}| < 1$,  \\ 
\frac{1}{3} \sinb[2]{\CHDIR}  B_\TS^2 & if $|\lambda_{\TS}| > 1$. 
 }
\label{equ:Fluctuation}
\end{equation}
Equations~\eref{equ:Drift} and~\eref{equ:Fluctuation} are the expectation and 
variance of a uniform distribution of $\lambda_{\TS + 1}$.

For small $\CHDIR > 0$, 
the trajectories of direction-sweep MCMC are exactly equivalent to those of a 
Gaussian process:
\begin{equation}
   \lambda_{\TS + 1} = \lambda_{\TS} + [\expecta(\lambda _{\TS+1}|\lambda 
   _{\TS}) - 
   \lambda_{\TS}] + \sqrt{\sigma^2(\lambda_{\TS + 1}|\lambda _{\TS})} \, w_t,
   \label{eq:ImpactSimulation}
\end{equation} 
where $w_\TS$ samples a standard normal distribution. 
For small $\CHDIR$, the fluctuation term $\sqrt{\sigma^2(\lambda_{\TS + 
1}|\lambda _{\TS})}$ is proportional to $\CHDIR$. The drift 
$\expecta(\lambda _{\TS+1}|\lambda_{\TS}) - \lambda_{\TS}$ is proportional to 
$(\CHDIR)^2$ for $|\lambda| > 1$ and proportional to $\CHDIR$ for $|\lambda| < 
1$. This leads to distinctive dynamics 
for $|\lambda|>1$ and for $|\lambda| < 1$, 

\subsection{Excursions ($|\lambda| > 1$)}
\label{sec:excursions}
For $|\lambda_{\TS}| > 1$, \eref{eq:ImpactSimulation} becomes 
\begin{equation}
   \lambda_{\TS + 1} = \lambda_{\TS} 
   - \frac{\lambda_\TS}{2} (\CHDIR)^2 
   + \sqrt{\frac{1}{3} B_\TS^2 (\CHDIR)^2}\,w_\TS
\end{equation}
in the limit of small $\CHDIR$. This equation agrees with the
discrete-time Langevin equation 
\begin{equation}
 \lambda_{\TS+1}  = \lambda_\TS + D^{(1)}(\lambda_\TS, t_\TS) \tau + 
 \sqrt{2D^{(2)}(\lambda_\TS, t_\TS) \tau} \, w_\TS,  
\end{equation}
where the discrete times $t_t$ are separated by the time step $\tau$, and 
$D^{(1)} 
$ and $D^{(2)} $ are the Kramers--Moyal expansion coefficients of the 
time-dependent probability distribution of $\lambda$ that correspond to drift 
and diffusion, respectively (see \cite[eq.~(3.138)]{Risken1989}).
For $|\lambda| > 1$, the \IMPACT thus performs  an \quot{excursion},  a random 
walk in the quantity $\tau = (\CHDIR)^2$. Such an 
excursion corresponds to a total number of time steps that scales 
as $\sim \const / (\CHDIR)^2$. As each time step increases the direction 
by $\CHDIR$, this increases the rolled-out direction by an amount that diverges 
as $ \sim  \const/ \CHDIR$. The excursions with $|\lambda| > 1$ for different 
$\CHDIR$ therefore become similar if the trajectory of the impact parameter is
recorded as a function of $(\RDIR_\TS - \RDIR_0)\CHDIR$ (see 
\subfig{fig:ThetaLambdaTrajectory}{a and b}).
 
During the excursion, $\xvec_\TS$ follows the rotation of the direction in 
steps of 
$\CHDIR$ (see \subfig{fig:ThetaLambdaTrajectory}{c}). Thus, the difference in 
the rolled-out dipole angle during an excursion 
diverges as $ \sim  \const / \CHDIR$. For $\CHDIR > 0$, 
this 
is evidenced by the unit slope of the increasing parts of the trajectories of 
$(\RDA_\TS - \RDA_0) \CHDIR$ as a function of $(\RDIR_\TS - \RDIR_0) \CHDIR$ 
(see \subfig{fig:ThetaLambdaTrajectory}{a and b}, again). 

\subsection{Zigzags ($|\lambda| < 1$)}
\label{sec:zigzags}
For $|\lambda_\TS| < 1$, the fluctuations in \eref{equ:Fluctuation} are
negligible for small $\CHDIR$ and 
the Gaussian process of 
\eref{eq:ImpactSimulation} approaches its deterministic limit:
\begin{equation}
   \Delta\lambda = \lambda_{\TS + 1} - \lambda_{\TS} = 
   \mp \frac{\sqrt{\RAT^2 - \lambda_{\TS}^2} + \sqrt{1 - \lambda_\TS^2}}{2} \, 
   \CHDIR,
\end{equation}
where the negative sign on the right-hand side is for $\xvec_\TS \in 
\SEG^+$, and the positive sign for $\xvec_\TS \in 
\SEG^-$. For small $\CHDIR$, this becomes  
a non-linear 
differential equation whose solution (up to an integration constant) is
\begin{equation}
   \DIR(\lambda) = \pm \frac{
   \arcsin(\lambda) 
   - \RAT^2\arcsin(\lambda / \RAT)
   + \lambda [C(\lambda) - B(\lambda)]}{\RAT^2 - 1},
\label{equ:ExactZigZag}   
\end{equation} 
where $B(\lambda)=\sqrt{\RAT^2 - \lambda^2}$ and 
$C(\lambda)=\sqrt{1-\lambda^2}$.
The numerical trajectories 
(for $|\lambda| <1$) reproduce the exact solution of \eref{equ:ExactZigZag} 
(see 
\subfig{fig:ThetaLambdaTrajectory}{c}). 

For small $\CHDIR > 0$, if $\xvec$ enters the segment $\SEG^+$ with unit
impact parameter $\lambda = 1$, $\lambda$ will decrease at every step until
$\lambda = -1$. In $\SEG^-$, the impact parameter likewise increases from
$\lambda = -1$ to $\lambda = 1$. This trapped deterministic motion creates one
\quot{zigzag} in the trajectory of $|\lambda|$. The total change of 
the rolled-out direction $\Delta\RDIR_\text{\ZIG}$ during such a zigzag is 
given by
\begin{equation}
   \Delta\RDIR_\text{\ZIG} = \frac{2\RAT^2 \mathrm{arccsc}(\RAT) + 
   2\sqrt{\RAT^2 - 1} - 
   \pi}{\RAT^2 - 1 },
\label{eq:ChangeSaw}
\end{equation}
and is always smaller than $\pi$. In a reference frame with $\phi_\TS=0$, 
(which is 
rotated by $-\CHDIR$ at every time step), $\xvec$ performs a negative rotation
(see the trajectory from $B$ to $D$ in
\subfig{fig:ThetaLambdaTrajectory}{c}), and thus follows the rotation of
the system. Therefore, the rolled-out dipole angle $\RDA_\TS$ remains roughly
constant, leading to a plateau in the non-rotating reference frame.
For $\lambda_\TS = 0$, the dipole angle $\RDA_{\TS}$ is independent of the
precise position $\xvec_\TS$ on its segment. At the center of each plateau,  
the fluctuations of $\RDA_{\TS}$ thus vanish even at finite $\CHDIR$ (see point 
$C$ in \subfig{fig:ThetaLambdaTrajectory}{c}).

Since the total change of the rolled-out direction $\Delta\RDIR_\text{\ZIG}$ in 
\eref{eq:ChangeSaw} is independent of $\CHDIR$, the zigzags in the trajectories 
of $|\lambda_\TS|$ for different (small) values of $\CHDIR$ are similar if 
plotted as a function of $\RDIR_\TS - \RDIR_0$, but ever steeper 
as a function of $(\RDIR_\TS - \RDIR_0)\CHDIR$ (see 
\subfig{fig:ThetaLambdaTrajectory}{a and b}).

\subsection{Interplay of excursions and zigzags}

\begin{figure}
\begin{indented}
\item[]
\includegraphics[scale=\figurescale]{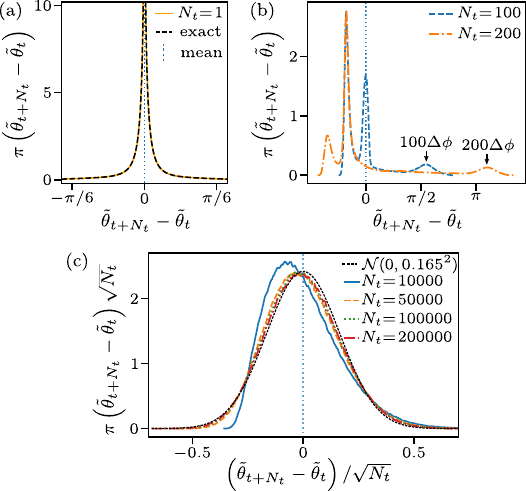}
\end{indented}
\caption{Distributions $\pi(\RDA_{\TS + N_t} - \RDA_\TS)$ 
   for $\RAT=1.1$ and $\CHDIR=\pi/180$. 
   \subcap{a} $N_t=1$. \subcap{b} Distributions for moderate $N_t$. 
   \subcap{c} Rescaled large-$N_t$ distributions compared to a Gaussian 
distribution.} 
\label{fig:SymmNonSymm}
\end{figure}

After one rapid motion from $\lambda =1$ to $\lambda = -1$ or vice versa, 
the trajectory may switch segments to 
continue at $| \lambda|  < 1$, adding one more leg to the 
negative-rotation zigzag of $\lambda_\TS$
(or two legs to the zigzag of $|\lambda_\TS|$ in 
\subfig{fig:ThetaLambdaTrajectory}{c}).
The trajectory may also 
switch to an excursion with $|\lambda| > 1$, that is,  
to a positive rotation of the rolled-out dipole 
angle $\RDA$.
The time (number of steps) of an excursion scales as $\sim \const / 
(\CHDIR)^2$ whereas the time of one zigzag is shorter by factor 
of $\CHDIR$ as it scales as 
$\sim \const / \CHDIR$. Nevertheless, positive and negative rotations balance, 
and the expectations $\expecta{(\RDA_{\TS + N_t} - \RDA_\TS)}$ are zero for all 
values of $N_t$. This holds for a single move
($N_t=1$)
because $\pi(\RDA_{\TS + 1} - \RDA_{\TS})$ is symmetric 
in consequence of the detailed balance 
of the move from $\TS$ to $\TS+1$ (see \subfig{fig:SymmNonSymm}{a}). 
For $N_t > 1$, $\RDA_{\TS + N_t} - \RDA_{\TS} = \RDA_{\TS+N_t} - 
\RDA_{\TS + N_t - 1} + \dots + \RDA_{\TS+1} - \RDA_\TS  $ yields
\begin{eqnarray}
\fl
   \expectb{\RDA_{\TS + N_t} - \RDA_\TS} = 
   \expectb{\RDA_{\TS + N_t} - \RDA_{\TS + N_t - 1}}   
   + \dots \nonumber\\
   + \expectb{\RDA_{\TS + 2} - \RDA_{\TS+1}} 
   + \expectb{\RDA_{\TS + 1} - \RDA_\TS} = 0, 
\label{equ:Telescopic}
\end{eqnarray}
because the expectation of a sum of (possibly dependent) random variables 
equals the sum of expectations. The distribution $\pi(\RDA_{\TS + N_t} - 
\RDA_{\TS})$, although it is of zero expectation,  can be highly asymmetric
(see \subfig{fig:SymmNonSymm}{b}). For $N_t  
\lesssim \const /(\CHDIR)^2$, the distribution peaks for large $\RDA$ that 
corresponds to trajectories that remain on long excursions. For $N_t \gg 
\const /(\CHDIR)^2$, the distribution approaches a Gaussian and becomes 
again symmetric because the large number of steps allows excursions and 
zigzags to compensate in a single trajectory (see \subfig{fig:SymmNonSymm}{c}).
The vanishing of $ \expecta{(\RDA_{\TS + N_t} - \RDA_\TS) }$ implies that 
there are \bigO{1/\CHDIR} zigzags for each excursion. 
Microscopically, this can be understood through the existence of the cutoff 
value $\LCUTP$ (see \eref{eq:lcut} and \subfig{fig:Sequential}{b}).
If $|\lambda_{\TS}| > \LCUTP$, the next value of the impact parameter
$|\lambda_{\TS+1}| > 1$; in contrast, a current value of the impact parameter 
$1 < |\lambda_\TS| < \LCUTP$ may either
produce $|\lambda_{\TS+1}| > 1$ or $|\lambda_{\TS+1}| < 1$. In the latter case,
$\xvec$ gets trapped in its corresponding segment.

\section{Approach to equilibrium}
\label{sec:convergence}

The trajectory of the dipole in its sample space $\Omega = \SET{(\DD, \DA)}$ is 
characterized by persistent negative and positive rotations. In order to 
quantify the approach to equilibrium of direction-sweep MCMC, we 
consider mixing times~\cite{Levin2008,Diaconis2011}
\begin{equation}
\label{eq:mix}
   \TMIX(\varepsilon) = \min\{\TS:d(\TS)\leq \varepsilon\},
\end{equation}
where $d(\TS)$ is the total variation distance (TVD) between the stationary 
distribution $\pi$ and the probability distribution
$P^\TS(\xvecbar_0, \cdot)$ at time $\TS$ obtained by starting from the most 
unfavorable lifted initial configuration $\xvecbar_0$:
\begin{eqnarray}
   d(t) &= \max_{\xvecbar_0} \, ||P^\TS(\xvecbar_0, \cdot) - \pi||_\text{TVD} 
   \\ 
   &= \max_{\xvecbar_0} \, \half \int \diff \xvecbar \, 
   |P^\TS(\xvecbar_0, \xvecbar) - \pi(\xvecbar)|.
\label{equ:TVD_def}   
\end{eqnarray}
The time $\TMIX = \TMIX(1/4)$ is defined as the mixing time.
For $\varepsilon < 1/4$, $\TMIX(\varepsilon) $ is bounded 
through $\TMIX$, 
\begin{equation}
	\TMIX(\varepsilon) \leq \lceil \log_2 \varepsilon^{-1}\rceil \, \TMIX,
\end{equation}
showing that the mixing process is exponential~\cite{Levin2008}. 

We checked numerically for the dipole that the same $\xvecbar_0$ maximizes the 
TVD for 
$\varepsilon$ in the neighborhood of $1/4$ and determine $\TMIX$ \emph{via}
the time $\TMIX(\xvecbar_0)$:
\begin{equation}
   \TMIX(\xvecbar_0)
   = \min \{ 
   \TS: ||P^\TS_\DA(\xvecbar_0, \cdot) - \pi_\DA||_\text{TVD} \leq 1/4 
   \}.
\end{equation}
The maximum of $\TMIX(\xvecbar_0)$ over the initial configurations $\xvecbar_0$ 
then yields $\TMIX$ (by doing so, we effectively interchanged the $\min$ in 
\eref{eq:mix} with 
the  $\max$ in \eref{equ:TVD_def}). Due to the rotational invariance of the 
ring 
system, 
$\TMIX(\xvecbar_0)$ only depends on the angle difference $\DA_0 - \DIR_0$. 
We thus set $\DIR_0=0$ and consider $\TMIX(\DD_0, \DA_0)$, which we 
determine numerically by running $100000$ simulations that all start from 
$\xvecbar_0 = (\DD_0, \DA_0)$. At each time step $\TS$, we use these 
simulations 
to determine $P^\TS_\DA(\xvecbar_0, \cdot)$ and its TVD with $\pi$. The 
evaluation of the TVD requires in our case the evaluation of a two-dimensional 
integral over $\Omega$. However, since $\DD$ relaxes very fast, we approximate 
it by the one-dimensional integral over $\DA$. 

\subsection{Identifying unfavorable initial configurations} 
\begin{figure}
\begin{indented}
\item[]
\includegraphics[scale=\figurescale]{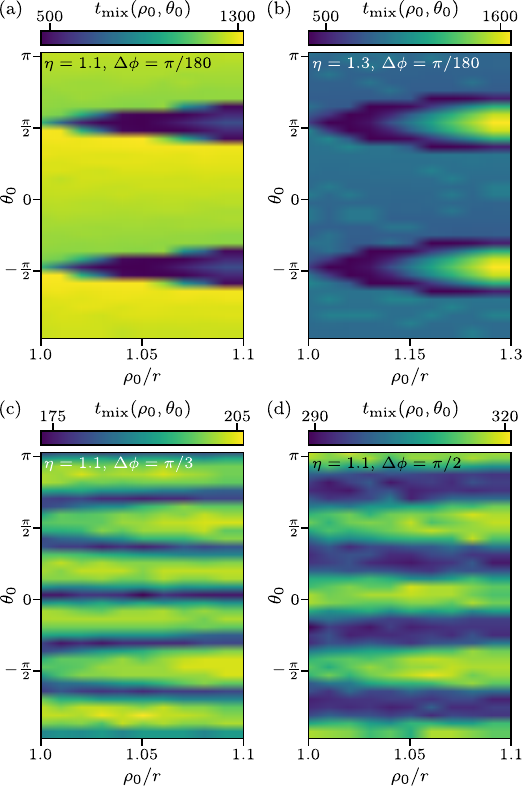}
\end{indented}
   \caption{Mixing times $\TMIX(\xvecbar_0 =(\DD_0, \DA_0))$ 
for different values of $\RAT$ and $\CHDIR$. }
\label{fig:heatmap}
\end{figure}

For small $\CHDIR > 0$, two unfavorable initial configurations stand out. First, 
trajectories with $\lambda_0 = 1$ and $\xvecbar_0 \in \SEG^+$ (or $\lambda_0 = 
-1$ and $\xvecbar_0 \in \SEG^-$) always start with a deterministic zigzag until 
$\lambda_{\TS}=-1$ (or $\lambda_{\TS} = 1$). At 
the time $t$ 
after this first zigzag, the 
probability distribution $P^\TS_\DA(\xvecbar_0,\cdot)$ is therefore strongly 
peaked and produces 
a large TVD in \eref{equ:TVD_def}. Thereafter, different trajectories either 
continue with more zigzags or else with an excursion, which then flattens 
$P^t_\DA(\xvecbar_0,\cdot)$. Second, a trajectory from $|\lambda_0|=\RAT$ 
starts with 
an excursion and $P^t_\DA(\xvecbar_0,\cdot)$ thus peaks at $\DA_\TS=\TS\CHDIR$. 
Once the 
random walk in $|\lambda_\TS|$ reaches $\LCUTP$, the distribution 
$P^\TS_\DA(\xvecbar_0,\cdot)$ starts to flatten. The most unfavorable initial 
configuration among these two depends on $\RAT$. For $\RAT\rightarrow 1$, we 
find  that starting the trajectory with a zigzag is the most unfavorable initial 
state, whereas starting the trajectory with an excursion is most unfavorable for 
larger $\RAT$ (see \subfig{fig:heatmap}{a and b}). This can be understood by 
the fact that the number of time steps in a zigzag increases as $\RAT \to 1$ 
(see \eref{eq:ChangeSaw}), whereas the difference $\RAT - \LCUTP$ that the 
impact 
parameter has to overcome in the initial excursion decreases (see 
\eref{eq:lcut}).

For large $\CHDIR$, the random walk of $\lambda_t$ is no longer described in 
terms of excursions and zigzags, and the initial configuration does not 
strongly influence $\TMIX(\xvecbar_0)$, except for values of $\CHDIR$ that 
correspond to small direction sets $\DCAL$. Then $\TMIX(\DD_0, \DA_0)$ is 
roughly periodic in $\DA_0$ (see \subfig{fig:heatmap}{c and d}). For 
$\CHDIR=\pi / 2$, this may be due to the fact that with $\DA_0\in\SET{-\pi/2, 
0, \pi/2, \pi}$ one of the two alternating directions hardly modifies 
$\theta_\TS$ during the initial part of the trajectory.

\subsection{Mixing Time}

We now systematically  study $\TMIX$ for direction-sweep MCMC as a 
function of the 
direction
set $\DCAL$. We compare  it with the 
reversible version that samples $\DIR_{t+1}$ randomly from $\DCAL$ (random 
discrete MCMC), and also 
with reversible MCMC with continuous directions  $\CHDIR = \mathrm{ran}(0, 
\pi)$ 
(random continuous MCMC). Both versions satisfy detailed balance. 

\begin{figure}
\begin{indented}
\item[]
\includegraphics[scale=\figurescale]{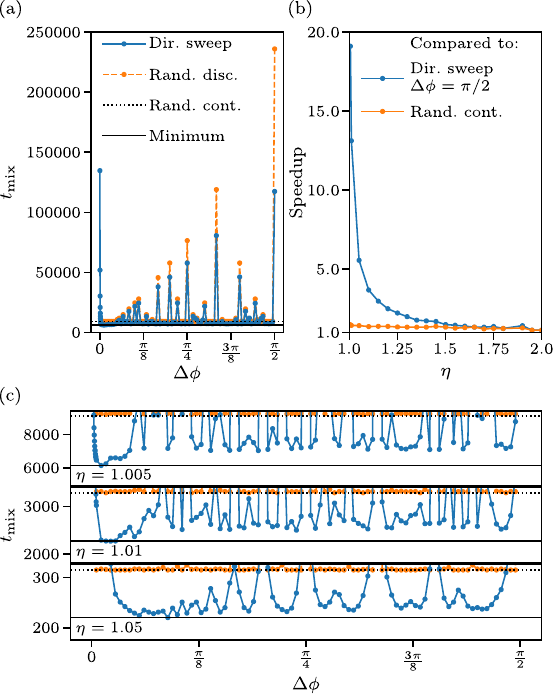}
\end{indented}
\caption{Mixing times $\TMIX$ of direction-sweep MCMC for different 
$\CHDIR$. 
   \subcap{a} $\TMIX$ at $ \RAT= 1.005$ compared to mixing times of random 
   discrete and continuous MCMC.
   \subcap{b} Optimal speedup with respect to $\CHDIR = \pi/2$ and to 
   random continuous MCMC. 
   \subcap{c} $\TMIX$ of direction-sweep MCMC compared to that of 
   random continuous MCMC for $\RAT \gtrsim 1$. The line labeling is the same 
   as in (a).}
\label{fig:TVD}
\end{figure}

Several properties stand out (see \subfig{fig:TVD}{a}). First, the 
mixing time is very sensitive to the size of $\DCAL$ regardless of whether its
elements are accessed sequentially or randomly. For a thin ring ($\RAT \to 
1$), 
the mixing time $\TMIX$ shows characteristic peaks for small direction sets 
$\DCAL$. The height of these peaks (for not too large set sizes $|\DCAL|$) is 
proportional to $1/|\DCAL|$. This yields a particularly large 
mixing time for $\CHDIR=\pi / 2$ where $|\DCAL| = 2$.

Second, we find that sweeping through the elements of $\DCAL$ is 
generically better than randomly sampling the direction from $\DCAL$, except  
for $\CHDIR \to 0$ where the sweeps are too slow and the mixing time 
diverges.
For small $|\DCAL|$, this benefit of 
direction-sweep MCMC is easily understood by the non-vanishing probability of 
repeated (redundant) moves in the same direction that only appear in random 
discrete MCMC. 

For all considered values of $\RAT$, we find that 
direction-sweep MCMC with appropriate $\CHDIR$ is faster than 
random continuous
MCMC and, in particular, as direction-sweep MCMC
with $\CHDIR = \pi / 2$
(see \subfig{fig:TVD}{b}). The speedup compared to 
the choice $\CHDIR=\pi/2$ is large, and it appears 
to diverge as $\RAT \to 1$. This may render the
non-reversible scheme especially promising for dipolar particles in ECMC 
where up to now $\CHDIR=\pi / 2$ was  always chosen.
We confirm that the smallest mixing time in direction-sweep MCMC is indeed 
reached for small $\CHDIR$, that is, for the peculiar trajectories 
discussed in \sref{sec:equilibrium}.
This, in our model, can of 
course only be 
observed for $\RAT \to 1$ because the speedup for small $\CHDIR$ is cut off by 
the divergence of $\TMIX$ for $\CHDIR \to 0$.

\begin{figure}
\begin{indented}
\item[]
\includegraphics[scale=\figurescale]{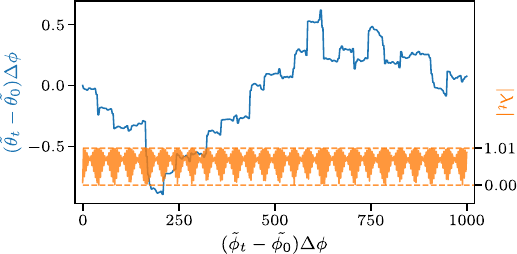}
\end{indented}
\caption{Trajectories of $(\RDA_\TS - \RDA_0)\CHDIR$ (blue, upper) 
   and of $|\lambda_\TS|$ (yellow, lower) \emph{vs.}
   $(\RDIR_\TS - \RDIR_0) \CHDIR $ for $\RAT= 1.01$ 
   and $\CHDIR = 85 \, \pi / 180$.}
\label{fig:LargeDeltaPhiTrajectory}
\end{figure}

Finally, we find that direction-sweep MCMC is usually faster than the random 
continuous MCMC even for large values of $\CHDIR$, except when $|\DCAL|$ is 
very 
small (see \subfig{fig:TVD}{c}, large angles $\CHDIR$ give small direction sets 
only if $\CHDIR / (2 \pi)$ is a simple fraction). The trajectories remain very 
regular  for generic $\CHDIR$. They feature intriguing patterns for the impact 
parameter $\lambda_\TS$ and the 
rolled-out dipole angle $\RDA_\TS$, that require further study (see 
\fref{fig:LargeDeltaPhiTrajectory}).

\section{Conclusion}
\label{sec:conclusion}

We have discussed a non-reversible MCMC algorithm for particle systems that 
sweeps through the direction of motion, rather than to sample directions 
randomly. For a single two-dimensional dipole, we proved in a local-equilibrium 
limit that direction-sweep 
MCMC induces 
persistent dipole rotations with a total rolled-out angle that 
diverges as the 
direction sweep becomes slower. Persistent rotation takes place 
in both senses, and the two exactly compensate to zero net rotation. 

Direction-lifted MCMC (of which direction-sweep MCMC is a special case) 
remains valid for general $N$-body problems. It preserves
the independence of the lifted stationary 
distribution from the lifting 
variable~\cite{Krauth2021eventchain} even if the thermalization 
condition at fixed lifting 
variable is dropped. 
Real-world direction-lifted MCMC may go to much smaller values of 
$\CHDIR$ than the single dipole, simply because mixing times will be much 
larger in applications. It will thus be fascinating to 
understand the usefulness of direction lifting
for applications such as polymer physics, and also in systems of 
long-range 
interacting extended molecules at the core of the JeLLyFysh 
project~\cite{Hoellmer2020}.
More generally, our model illustrates 
that non-reversibility profoundly changes the basic 
properties of 
local MCMC algorithms, in the same way as out-of-equilibrium statistical 
physics is fundamentally different from its equilibrium counterpart.

\ack

W.K. acknowledges support from the Alexander von Humboldt Foundation. We thank 
A. C. Maggs for helpful discussions.

\section*{References}
\providecommand{\newblock}{}


\begin{thebibliography}{10}
\expandafter\ifx\csname url\endcsname\relax
  \def\url#1{{\tt #1}}\fi
\expandafter\ifx\csname urlprefix\endcsname\relax\def\urlprefix{URL }\fi
\providecommand{\eprint}[2][]{\url{#2}}

\bibitem{Metropolis1953JCP}
Metropolis N, Rosenbluth A~W, Rosenbluth M~N, Teller A~H and Teller E 1953 {\em
  J. Chem. Phys.\/} {\bf 21} 1087--1092

\bibitem{Levin2008}
Levin D~A, Peres Y and Wilmer E~L 2008 {\em {Markov Chains and Mixing Times}\/}
  (American Mathematical Society)

\bibitem{OKeeffe2009}
O'Keeffe C~J and Orkoulas G 2009 {\em J. Chem. Phys.\/} {\bf 130} 134109

\bibitem{KapferKrauth2017}
Kapfer S~C and Krauth W 2017 {\em Phys. Rev. Lett.\/} {\bf 119}(24) 240603

\bibitem{Lei2018b}
Lei Z and Krauth W 2018 {\em EPL\/} {\bf 124} 20003

\bibitem{RenOkeeffe2007}
Ren R, O'Keeffe C~J and Orkoulas G 2007 {\em Mol. Phys.\/} {\bf 105} 231--238

\bibitem{Berg2004book}
Berg B~A 2004 {\em {Markov Chain Monte Carlo simulations and their statistical
  analysis: with web-based Fortran code}\/} (World Scientific) ISBN
  9789812389350

\bibitem{Diaconis2000}
Diaconis P, Holmes S and Neal R~M 2000 {\em Ann. Appl. Probab.\/} {\bf 10}
  726--752

\bibitem{SuwaTodoPRL2010}
Suwa H and Todo S 2010 {\em Phys. Rev. Lett.\/} {\bf 105}(12) 120603

\bibitem{Turitsyn2011}
Turitsyn K~S, Chertkov M and Vucelja M 2011 {\em Physica D\/} {\bf 240} 410 --
  414 ISSN 0167-2789

\bibitem{FernandesWeigelCPC2011}
Fernandes H~C and Weigel M 2011 {\em Comput. Phys. Commun.\/} {\bf 182}
  1856--1859 ISSN 0010-4655

\bibitem{BierkensPDMC2017}
Bierkens J, Bouchard-C{\^o}t{\'e} A, Doucet A, Duncan A~B, Fearnhead P, Lienart
  T, Roberts G and Vollmer S~J 2018 {\em Stat. Probab. Lett.\/} {\bf 136}
  148{\textendash}154

\bibitem{Bierkens2017}
Bierkens J and Roberts G 2017 {\em Ann. Appl. Probab.\/} {\bf 27} 846--882

\bibitem{Bernard2009}
Bernard E~P, Krauth W and Wilson D~B 2009 {\em Phys. Rev. E\/} {\bf 80}(5)
  056704

\bibitem{Michel2014JCP}
{Michel} M, {Kapfer} S~C and {Krauth} W 2014 {\em J. Chem. Phys.\/} {\bf 140}
  054116

\bibitem{Krauth2021eventchain}
Krauth W 2021 {\em Frontiers in Physics\/} {\bf 9} 229

\bibitem{Klement2019}
Klement M and Engel M 2019 {\em J. Chem. Phys.\/} {\bf 150} 174108

\bibitem{Michel2020}
Michel M, Durmus A and S\'en\'ecal S 2020 {\em J. Comput. Graph. Stat.\/} {\bf
  29} 689--702

\bibitem{Weigel2018}
Weigel R~F~B {Equilibration of Orientational Order in Hard Disks via Arcuate
  Event-Chain Monte Carlo} {Master thesis, Friedrich-Alexander-Universit\"at
  Erlangen-N\"urnberg, 2018}
  \urlprefix\url{https://theorie1.physik.uni-erlangen.de/research/theses/2018-ma-roweigel.html}

\bibitem{Faulkner2018}
Faulkner M~F, Qin L, Maggs A~C and Krauth W 2018 {\em J. Chem. Phys.\/} {\bf
  149} 064113

\bibitem{Hoellmer2020}
H\"{o}llmer P, Qin L, Faulkner M~F, Maggs A~C and Krauth W 2020 {\em Comput.
  Phys. Commun.\/} {\bf 253} 107168

\bibitem{WuTepperVoth2006}
Wu Y, Tepper H~L and Voth G~A 2006 {\em J. Chem. Phys.\/} {\bf 124} 024503

\bibitem{Mueller2020}
M\"{u}ller D, Kampmann T~A and Kierfeld J 2020 {\em Scientific Reports\/} {\bf
  10}

\bibitem{Kampmann2021}
Kampmann T~A, M\"{u}ller D, Weise L~P, Vorsmann C~F and Kierfeld J 2021 {\em
  Frontiers in Physics\/} {\bf 9} 635886

\bibitem{SMAC}
Krauth W 2006 {\em {Statistical Mechanics: Algorithms and Computations}\/}
  (Oxford University Press)

\bibitem{Chen1999}
Chen F, Lovász L and Pak I 1999 {\em Proceedings of the 17th Annual ACM
  Symposium on Theory of Computing\/}  275

\bibitem{Risken1989}
Risken H 1989 {\em {The Fokker-Planck Equation}\/} (Springer Berlin Heidelberg)

\bibitem{Diaconis2011}
Diaconis P 2011 {\em J. Stat. Phys.\/} {\bf 144} 445 ISSN 1572-9613

\end{thebibliography}
\end{document}